# Uniform large-area surface patterning achieved by metal dewetting for the top-down fabrication of GaN nanowire ensembles

**Jingxuan Kang, Rose-Mary Jose, Miriam Oliva, Thomas Auzelle, Mikel Gómez Ruiz, Abbes Tahraoui, Jonas Lähnemann, Oliver Brandt, Lutz Geelhaar**

*Paul-Drude-Institut für Festkörperelektronik, Leibniz-Institut im Forschungsverbund Berlin e.V., Hausvogteiplatz 5 – 7, 10117 Berlin, Germany*

E-mail: geelhaar@pdi-berlin.de

**Abstract**

The dewetting of thin Pt films on different surfaces is investigated as a means to provide the patterning for the top-down fabrication of GaN nanowire ensembles. The transformation from a thin film to an ensemble of nanoislands upon annealing proceeds in good agreement with the void growth model. With increasing annealing duration, the size and shape uniformity of the nanoislands improves. This improvement speeds up for higher annealing temperature. After an optimum annealing duration, the size uniformity deteriorates due to the coalescence of neighboring islands. By changing the Pt film thickness, the nanoisland diameter and density can be quantitatively controlled in a way predicted by a simple thermodynamic model. We demonstrate the uniformity of the nanoisland ensembles for an area larger than 1 cm². GaN nanowires are fabricated by a sequence of dry and wet etching steps, and these nanowires inherit the diameters and density of the Pt nanoisland ensemble used as a mask. Our study achieves advancements in size uniformity and range of obtainable diameters compared to previous works. This simple, economical, and scalable approach to the top-down fabrication of nanowires is useful for applications requiring large and uniform nanowire ensembles with controllable dimensions.

## 1. Introduction

Semiconductor nanowires are the subject of intense research, not least because of the conceptual advantages that they offer for various applications compared to thin films.[1] Many devices benefit from large arrays of ideally identical nanowires, in particular light-emitting diodes (LEDs), solar cells, and photoelectrochemical cells.[2–4] The different approaches to the synthesis of nanowires can generally be divided into bottom-up and top-down ones. In the former ones, the nanowires are grown from the vapor or liquid phase, and in the latter ones, the nanowires are obtained by the structuring of a bulk crystal or thin film. Both approaches have principal advantages and disadvantages.[5] For example, doping is typically difficult to control for bottom-up synthesis because the growth mechanisms are very complex, whereas in top-down fabrication the nanowire doping is simply inherited from the mature doping technology established for thin film and bulk growth. The surface patterning needed to control which material is not removed in top-down processing can be obtained by different techniques,[6] and the same techniques can also be used to guide bottom-up growth and determine the nanowire position and diameter.[7] Possibly, most popular in academic research is the use of electron beam lithography due to its high spatial resolution and flexibility, but this method does not allow the patterning of large areas with high throughput.

Arguably, the simplest approach to surface patterning is the dewetting of a metal film.[8–10] Dewetting is a phenomenon where a liquid or solid film on a solid surface breaks up into three-dimensional (3D) islands to minimize the free energy.[11,12] For dewetting to take place in solid thin films, mass transport needs to be activated, typically by annealing. The resulting island ensemble can be used not only as a mask for top-down nanowire fabrication, but also as seeds for bottom-up growth. The major advantages of this approach to surface patterning are that the process is simple, can be scaled to large areas, and does not require any kind of lithography equipment. On the downside, like any self-assembly process, dewetting exhibits limitations regarding the level of control over feature size and ensemble homogeneity. These aspects have not been addressed in most publications on dewetting-based nanowire synthesis, and even in reports that study the dewetting process in more detail, little attention is paid to the optimization of the island size homogeneity and the predictability of its average value.[8–10,13–17] Thus, the full potential of metal dewetting for the synthesis of uniform nanowire arrays has not yet been realized.

Here, we investigate in detail the dewetting of Pt thin films for the top-down fabrication of GaN nanowire ensembles, focusing on the size and shape homogeneity including variations over large sample areas. Nanowires made of GaN and related alloys are a very promising materials platform for optoelectronic applications such as LEDs, lasers, single-photon emitters, and photoelectrocatalysis.[18] We choose Pt as the dewetting metal because it is less prone to chemically interact with GaN than other common choices such as Ni and Al[19,20] and it enables a higher island density than Au for the same film thickness.[10] We explore the influence of different annealing temperatures and durations, film thicknesses as well as underlying materials to identify the optimum conditions. Thus, we show how diameter and density can be selected on demand, obtain high uniformities for a wide range of diameters and number densities, and demonstrate homogeneity across a sample area of more than 1 cm².

## 2. Experiments

The overall process flow of the GaN nanowire top-down fabrication is illustrated in figure 1. The nanowires are fabricated from a commercial GaN(0001) template on sapphire that is initially covered with a buffer layer ($SiO_x$ or $SiN_x$) and a Pt film. In the next step, the sample is annealed and the Pt film breaks up into an ensemble of nanoislands. These islands serve as a mask for the nanopatterning of the GaN following previously reported etching steps.[21,22] First, the buffer layer that is not protected by the nanoislands is removed by reactive ion etching (RIE). In the next step, the exposed GaN is dry-etched by inductively coupled plasma (ICP) reactive ion etching. An additional KOH wet etching removes surface defects induced by the dry etching and ensures smooth vertical nanowire sidewalls. In the end, the buffer layer and the Pt islands on top are removed by an HF bath.

The focus of the current study is the Pt dewetting step since ideally it controls the size, shape, and spatial distribution of the final nanowires. Pt films of thicknesses in the range (3–15) nm were deposited by electron beam evaporation at a rate of 0.2 nm/s on GaN, $SiO_x$, and $SiN_x$ surfaces. The samples were annealed either in $N_2$ in a rapid thermal annealing (RTA) furnace for 5 min at temperatures in the range of (600–850) °C or in an Ar-filled tube furnace for 0 min (only ramp-up and -down) to 3 h at temperatures in the range of (800–1000) °C. The temperature in the tube furnace was ramped up at a rate of 25 °C/min and ramped down by natural cooling (heat transfer to the lab environment).

After the formation of the nanoislands through metal dewetting, the samples underwent a series of sequential etching procedures to fabricate GaN nanowires. The first step was reactive-ion etching (Sentech SI 500), where the $SiO_x$ layer exposed between the Pt nanoislands was removed using a $CF_3$ plasma with an RF bias of 250 V and a pressure of 3 Pa. Subsequently, the samples were anisotropically etched via inductively coupled plasma (ICP) etching (SAMCO RIE140iP) at a pressure of 1.3 Pa with a $BCl_3:Cl_2$ gas ratio of 20:5, an RF power of 25 W, and an ICP power of 100 W. Following the dry etching of $SiO_x$ and GaN, the resulting GaN nanowires were immersed in a heated KOH solution with a 30% mass fraction for a duration of 150 s. The temperature of the KOH solution was maintained between 60 °C and 70 °C throughout this process.

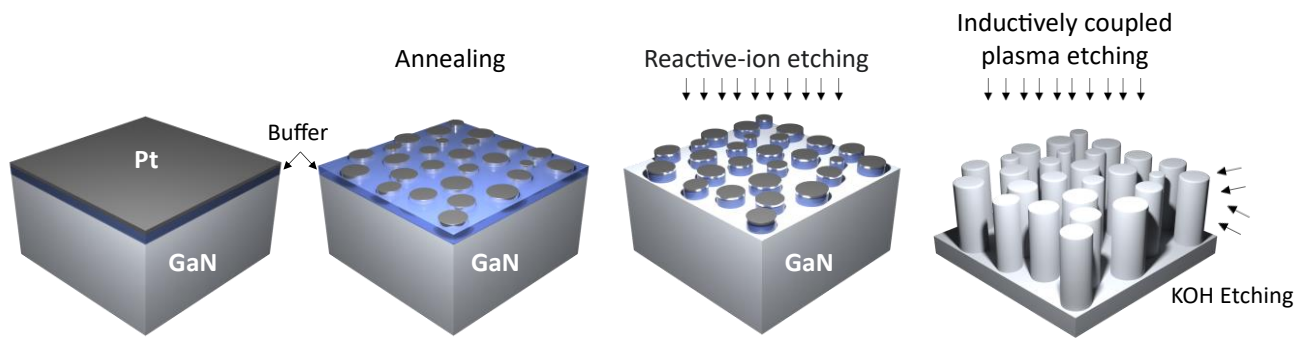

Figure 1: Process flow for the top-down GaN nanowire fabrication using a Pt dewetting mask.

After annealing, the surface morphology of the samples was characterized by either atomic force microscopy (AFM) or scanning electron microscopy (SEM). Statistical information on the nanoisland size, shape, and number density was extracted from the SEM data by image processing using the software ImageJ and a Python code developed by us. This procedure is described in more detail in the supporting information (figure S1).

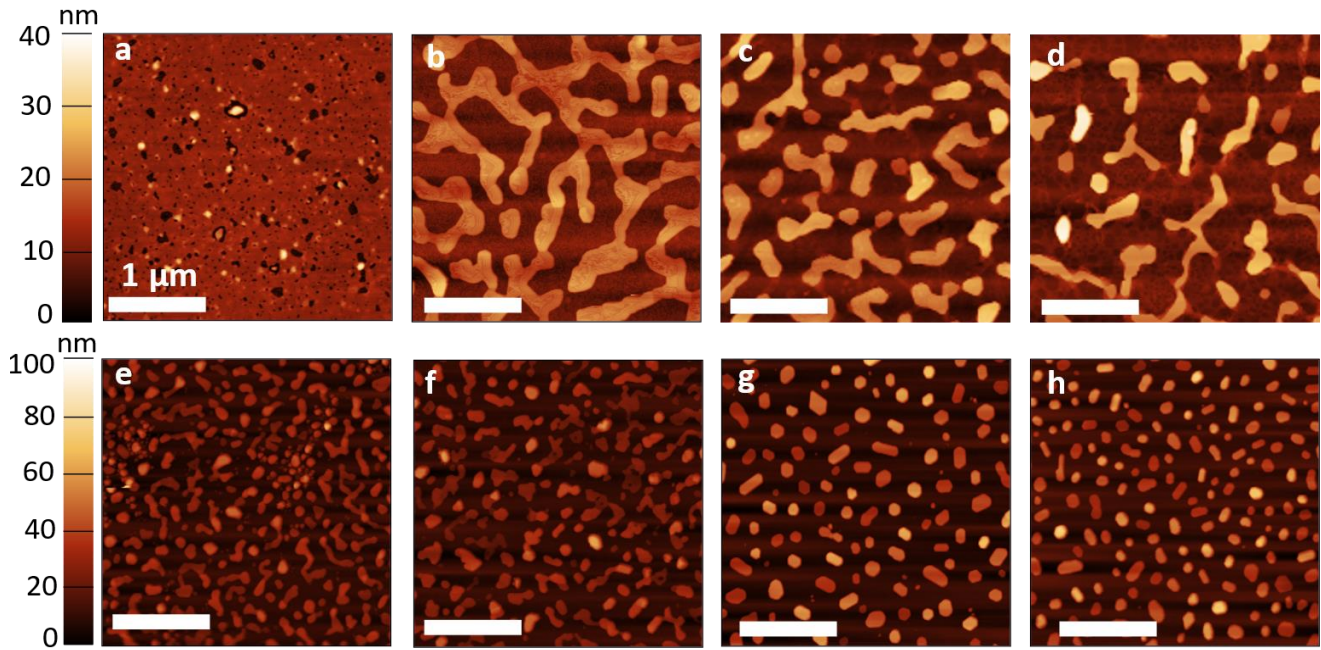

Figure 2: AFM topographs of Pt(8 nm)/GaN samples annealed in the RTA furnace for 5 min at a temperature of (a) 600 °C, (b) 700 °C, (c) 800 °C, (d) 850 °C, and of Pt(5 nm)/GaN samples annealed in the tube furnace at 800 °C for (e) 0 min, (f) 20 min, (g) 40 min, and (h) 60 min. Topographs (a)–(d) and (e)–(h) share the same height scale, respectively.

## 3. Results and discussion

In order to analyze the initial evolution of the Pt layer during annealing, we consider data for moderate temperatures and a short duration obtained with the RTA furnace. Figures 2(a)–(d) shows the morphology of Pt films with 8 nm thickness directly deposited on GaN(0001) ['Pt(8 nm)/GaN' samples in the following] annealed for 5 min at different temperatures. As seen in figure 2(a), at the lowest temperature of 600 °C voids form in the continuous Pt film, indicating the onset of the dewetting. As the annealing temperature increases, as displayed in figures 2(b)–(d), the voids expand, leading to the formation of finger-shaped islands, and then these islands break up into smaller islands. Similar results were reported for Pt films deposited on $SiO_2$ and annealed under comparable conditions.[13]

The influence of the annealing duration is studied for Pt(5 nm)/GaN samples annealed for 0–60 min at 800 °C in the tube furnace. The corresponding AFM topographs are depicted in figures 2(e)–(h). Simple ramp-up and –down of the temperature [duration 0 min, panel (e)] results in finger-shaped islands. This morphology resembles the one in figure 2(d) obtained for RTA of 5 min duration at 850 °C but with smaller islands, suggesting an overall similar but somewhat higher temperature budget in comparison to the different amount of Pt. With increasing annealing duration [panels (f)–(h)], the island shapes become more roundish and regular, their height increases, and the separations between them widen. For a combination of 800 °C and 1 h, the islands exhibit a fairly circular shape and high size homogeneity.

The Pt dewetting process investigated in these experiments is consistent with the void growth model.[11,23] According to this model, voids open up in a continuous metal film when a thermal treatment provides sufficient energy to activate mass migration. Capillary forces cause the void edges to deform into fingers and the film breaks up into 3D islands. These 3D islands tend to form a circular shape in order to minimise their free energy.

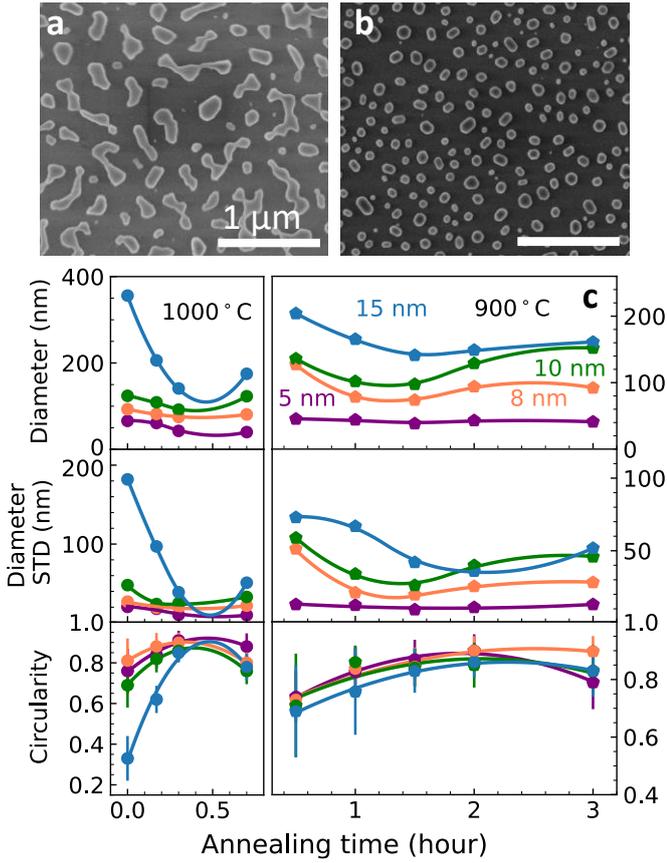

Figure 3: SEM images of Pt(10 nm)/SiO$_x$(220 nm)/GaN samples annealed at 900 °C for (a) 30 min and (b) 1.5 h. (c) Mean equivalent disc diameter, diameter standard deviation (STD), and mean circularity of nanoisland ensembles as a function of annealing time for different Pt film thicknesses on SiO$_x$/GaN annealed at 900 °C and 1000 °C in the tube furnace. The corresponding diameter and circularity histograms are included in the supporting information.

For most nanowire applications, a high size uniformity of the nanowire ensemble is important. We learn from our experiments and the void-opening model that with increasing thermal budget, the uniformity of the nanoisland size can be improved and their shapes resemble more and more circles as desirable for nanowire fabrication. In the following, we explore for different Pt film thicknesses what annealing conditions are needed in this respect. Since we expect high temperatures and long durations to be necessary, we choose to perform experiments with the tube furnace. Furthermore, we focus on Pt/SiO$_x$/GaN samples because these are most relevant for nanowire fabrication, as explained in a later section.

The general trend that we find is illustrated by the two secondary electron (SE) micrographs in figure 3(a) and (b) that correspond to 10 nm Pt film thickness, 900 °C annealing temperature, and durations of 0.5 and 1.5 h, respectively. For the shorter duration, mostly irregular islands with a wide range of shapes and sizes are obtained. For the longer duration, all of the islands are fairly circular and the size distribution has narrowed. This evolution is consistent with the findings for Pt directly on GaN and the void-opening model discussed above.

In order to quantitatively assess the island ensemble size and shape homogeneity, we extracted from such micrographs the mean equivalent disc diameter, the diameter standard deviation, and the mean circularity, as described in detail in the supporting information (figures S2 and S3). For each sample, 60–2000 nanoislands were considered, depending on nanoisland density. The results are presented in figure 3(c) as a function of annealing duration for different Pt film thicknesses (5–15 nm) and two annealing temperatures of 900 and 1000 °C. In general, with increasing time, initially the diameter and its standard deviation decrease and the circularity increases, indicating that the uniformity of the nanoislands improves. However, after an optimal time $t_o$ the uniformity tends to decrease again. $t_o$ varies with temperature and to some extent with Pt film thickness. Specifically, at 900 °C, the best island uniformity is reached after 1.5 h, even though for the 15-nm samples there is still a slight improvement for 2 h. At 1000 °C, $t_o$ is substantially reduced to 20 min.

These findings can be understood in the following way. The dewetting process and the evolution of the island ensembles require mass transport that is thermally driven. At the higher temperature of 1000 °C, this transport is faster, and $t_o$ decreases. For larger initial film thicknesses, more Pt needs to be transported, and thus $t_o$ increases. How the uniformity decreases for annealing beyond $t_o$ is illustrated in the supporting information (figure S4). For very long annealing times, islands much larger than the ensemble average form, and the area around these large islands is depleted of any other islands. This evolution can be considered as a consequence of coarsening.[24,25]

From a practical point of view, it is important that a careful optimization of annealing temperature and duration is needed to obtain highly uniform island ensembles. Comparing the smallest relative standard deviations of the diameters extracted for different Pt film thicknesses at temperatures of 900 °C and 1000 °C, we observe that the annealing at 900 °C results in slightly better uniformity, with relative standard deviations of approximately 24% (histograms are shown in the supporting information, figures S2 and S3). This value is slightly lower than the 28% reported previously for GaN nanowires fabricated through dewetting.[10] Also, the mean diameters in our study range from 40 nm to 150 nm, which is larger compared to the cited work with a range of 60 nm to 110 nm.

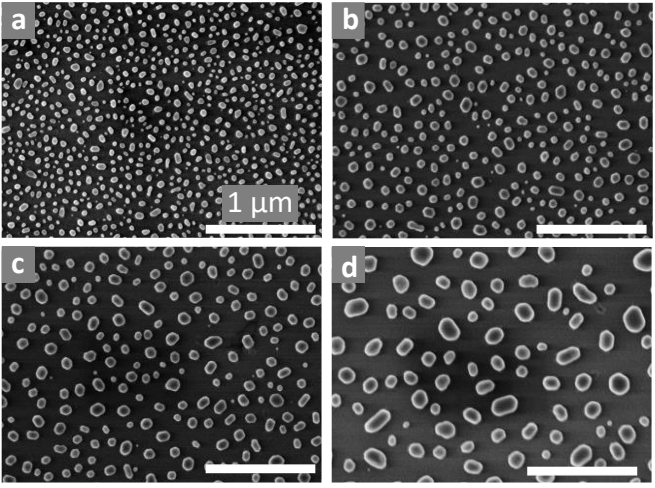

Figure 4: SEM images of Pt/SiO$_x$(220 nm)/GaN samples annealed at 900 °C for 1.5 h with film thickness of (a) 5 nm, (b) 8 nm, (c) 10 nm, and (d) 15 nm.

For the fabrication of nanowire ensembles, control over the average diameter is highly desirable. A substantial variation of the diameter with Pt film thickness is seen in the data of figure 3(c). This dependence is exemplified by the SE micrographs in figure 4 that were obtained for the Pt/SiO$_x$(220 nm)/GaN(0001) samples annealed for 1.5 h at 900 °C. We find that for this temperature, this duration gives the best overall uniformity. All the four micrographs show ensembles of nanoislands with fairly homogeneous size and roundish shape. The pronounced increase of the average diameter with Pt film thickness is accompanied by a decrease in number density.

Now, we analyze quantitatively the dependence of diameter and density on film thickness. Analytical formulas can be derived within a simple thermodynamic model that considers the free energies of the initial film and a final ensemble of identical, evenly spaced spherical-cap shaped islands.[11] This model ignores the formation mechanism as well all kinetic effects and takes into account temperature only indirectly via the surface free energies. Thus, it can provide only the minimum island diameter $d_{min}$. The dependences of this parameter and the island density $\rho$ on film thickness $h_A$ of the dewetting material A deposited on a surface of material B are given by:

$$d_{min} = 3Kh_A \quad (1)$$

$$\rho = \frac{4}{9K^3\pi g(\theta)} h_A^{-2} \quad (2)$$

with $K = \frac{2f(\theta)\gamma_A + \gamma_{AB} - \gamma_B}{g(\theta)(\gamma_A - \gamma_B + \gamma_{AB})}$, $f(\theta) = \frac{1+\cos\theta}{\sin^2\theta}$, and $g(\theta) = \frac{1+\cos\theta}{2}\sin^3\theta[1 + \cos\theta + \sin^2\theta]$. $\theta$ is the nanoisland contact angle, and $\gamma_i$ with $i$ = A, B, AB are the surface energies of materials A and B as well as the interfacial energy. Apart from dependencies on material parameters and geometrical considerations related to the spherical-cap shape, the minimum island diameter is proportional to the initial film thickness, and the density is inversely proportional to the square of the film thickness.

Figure 5 presents our experimental data for Pt films deposited on SiO$_x$, SiN$_x$, and GaN(0001) surfaces along with fits according to the above model. Since the model describes an influence of the surface energy, we consider three materials as substrate which have different surface energies.[26,27] In order to avoid the thermal decomposition of unprotected GaN, those samples were annealed at 800 °C for 1 h. For the other two surfaces, the data for 900 °C and 1.5 h were chosen, i. e., the conditions leading to best homogeneity and smallest diameter at a roughly similar temperature. For fitting the data, the parameters $K$ and $g(\theta)$ were freely adjusted.

In general, the experimental data points are described very well by the fits, implying that specific combinations of diameter and density can be fabricated on demand. Both parameters can be varied by about one order of magnitude even for the fairly small range of film thicknesses considered here. The good agreement between experiments and fits suggests that parameters outside the demonstrated range can be achieved in a straightforward way. We emphasize that diameters as small as 40 nm were obtained, and the density translates into average nearest-neighbor distances of similar magnitude, i. e., combinations that are attractive or applications requiring dense ensembles of thin nanowires. Furthermore, the data points and curves for GaN are clearly distinct from those for SiO$_x$ and SiN$_x$. In the thermodynamic model, the relevant material parameters are the surface and interface energies. Thus, the material differences observed here, in particular the notably higher slope for GaN in figure 5(a), are attributed to differences in surface and interface energies. The similarity between the data for SiO$_x$ and SiN$_x$ can be explained by the fact that both materials are amorphous, whereas GaN is single crystalline.

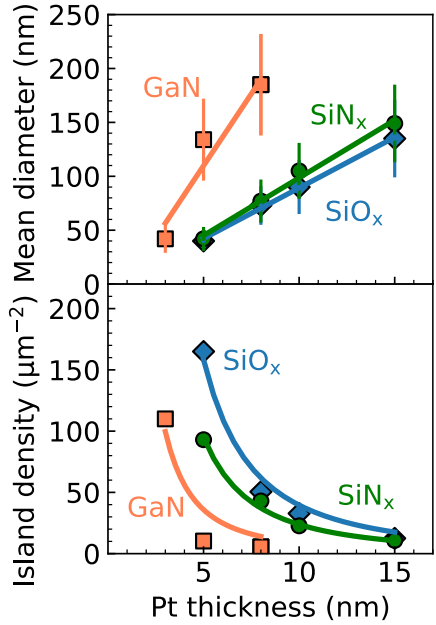

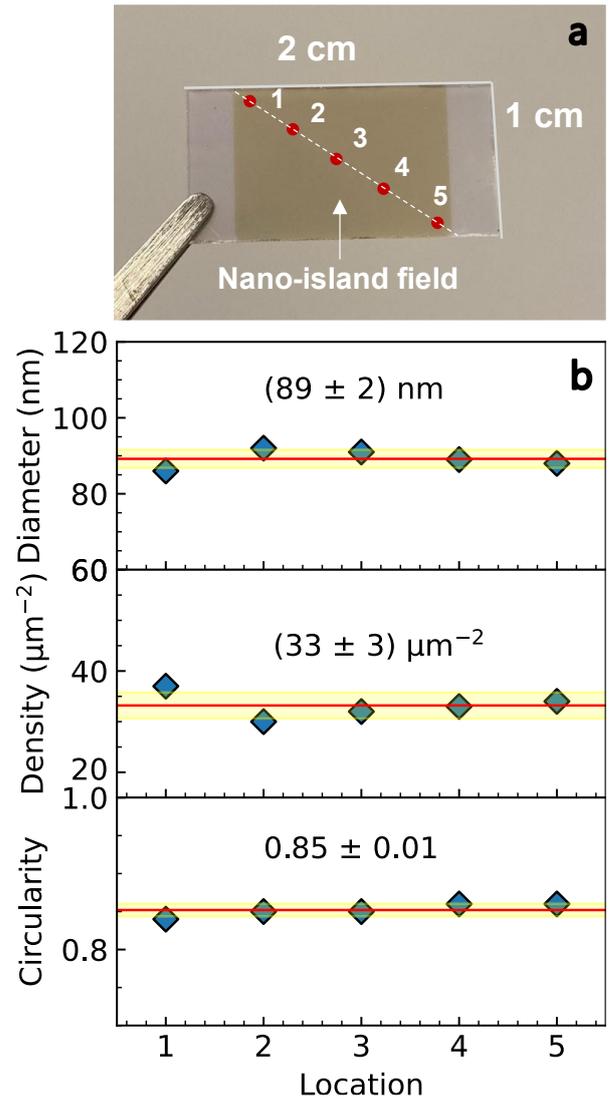

Figure 5: Dependence of nanoisland diameter and density on Pt film thickness for different dewetting substrates indicated as text labels. Data points correspond to annealing experiments and the solid lines are the result of fits as described in the main text.

A key advantage of using metal dewetting for the fabrication of nanowire ensembles is the scalability of this approach. In order to verify the uniformity of the nanoisland ensemble over a large area, we analyze a Pt dewetting pattern of size 1 cm × 1.5 cm as shown in figure 6(a). Five locations across the sample were characterised by SEM, and island equivalent disc diameter, number density, and circularity were extracted as described before. The results are presented in figure 6(b). The circularities are fairly close to 1, which is consistent with the results shown in figure 3. On the basis of the discussion of that figure, the data in figure 6(b) indicate very good local ensemble homogeneities. Between the five locations, the mean values are very similar. This finding is illustrated by the horizontal lines and shaded areas corresponding to the average over the five mean values and its standard deviation. This analysis demonstrates a uniform distribution of the islands over the entire sample surface.

Table 1: Comparison of island/nanowire density and diameter before and after etching, as determined for the SEM images depicted in figure 7(a) and (b).

| Etching | Density ($\mu m^{-2}$) | Mean diameter (nm) |
|---|---|---|
| Before | 139 | 45 |
| After | 119 | 44 |

Figure 6: Analysis of the island uniformity across a large Pt(10 nm)/SiO$_x$(220 nm)/GaN sample annealed for 90 min at 900 °C. (a) Photograph of the sample. The numbers indicate the locations at which SE micrographs were acquired. (b) Nanoisland equivalent disc diameter, density, and circularity determined at the five locations across the sample. The horizontal lines and shaded areas represent the average and standard deviation over the five locations, respectively.

The final step in our process development is the verification that the morphology of the nanoisland ensemble is transferred to the morphology of the nanowire ensemble. Figures 7(a) and (b) depict SE micrographs of a Pt(5 nm)/SiO$_x$/GaN sample before and after the etching process. Obviously, the morphologies are very similar. For a quantitative analysis, table 1 compares the diameter and the density of the nanoislands before the etching process and of the nanowires after the fabrication process. The diameters are almost identical, but the density is slightly lower after etching. This small discrepancy can be explained by the fact that thin neighboring nanowires tend to bundle during the SEM measurement due to charging, as the red circles indicate in figure 7(b). The resulting merged objects are counted as one nanowire during image analysis. Hence, we can conclude that the shape of the nanoisland mask is fully transferred to the nanowires. Therefore, the range of nanoisland diameters and densities as well as the corresponding uniformities that we showed above is accessible also for nanowire fabrication.

Ideally, the length of the nanowires can be controlled by the etching time. However, if the mask layer is too thin and/or soft, limitations are imposed with respect to the nanowire length. In particular, in preliminary experiments we found that the Pt islands by themselves are not robust enough. Thus, a sufficiently thick $SiN_x$ or $SiO_x$ buffer layer is needed, and in the etching optimization we focused on the latter. An additional benefit of the buffer layer is that it prevents the diffusion of Pt into the GaN during annealing, as demonstrated by energy dispersive X-ray spectroscopy (EDX) measurements in the supporting information. With a 220-nm-thick $SiO_x$ buffer layer, we obtained GaN nanowire lengths of up to 1 µm without deterioration of the ensemble uniformity.

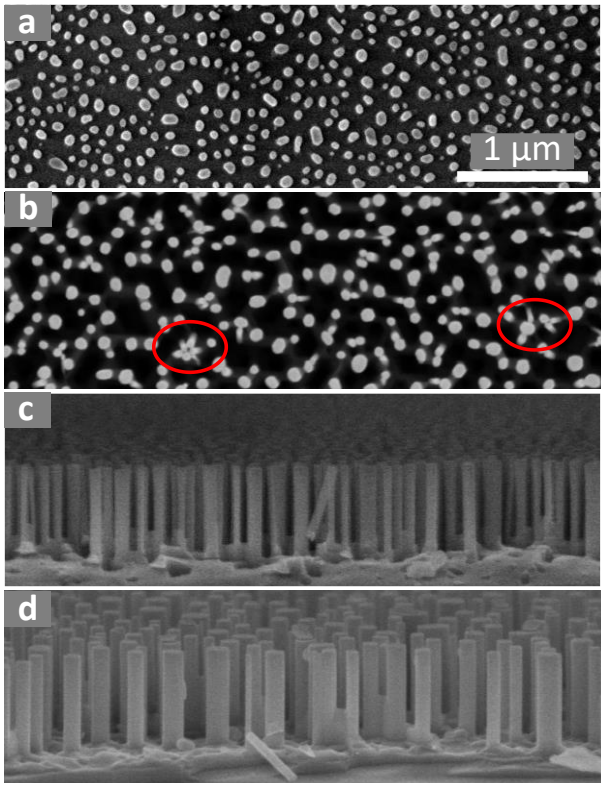

Figure 7: Top-view SE micrographs of a Pt(5 nm)/$SiO_x$(220 nm)/GaN sample (a) before and (b) after etching. The red circles in (b) indicate nanowire bundles formed during SEM observation. Side-view SEM images of GaN nanowires with mean diameters of (c) 40 nm and (d) 80 nm, resulting from Pt layers with initial thicknesses of 5 and 8 nm, respectively.

The morphology along the length of the nanowires is illustrated by the side-view SE micrographs in figures 7(c) and (d). The nanowire mean diameter is about 40 nm and 80 nm in figure 7(c) and (d), respectively. The sidewalls are smooth, and the diameter is homogeneous along the nanowires. This beneficial morphology is a consequence of the final KOH wet etch, that is known to remove the inclined sidewalls and defects induced by dry etching that are characteristic for the ICP etching.[21,28,29]

## 4. Summary and conclusion

We have studied in detail the dewetting of Pt thin films on different surfaces as a basis for the top-down fabrication of GaN nanowires. The dewetting process follows the void growth model for nanoisland formation. With increasing annealing duration, the nanoisland diameter uniformity and circularity improve but worsen at a certain point due to the coalescence of neighboring nanoislands. By carefully optimizing the annealing conditions, we have obtained uniform nanoisland ensembles with relative standard deviations of the diameter below 25%. The dependence of the diameter and density on Pt film thickness is quantitatively described by a simple thermodynamic model, thus allowing the predictive choice of experimental conditions for a desired morphology. Thus, we have demonstrated diameters in the range 40–150 nm and densities in the range 10–165 µm$^{-2}$. Even across a sample area of more than 1 cm², the average diameter varies by less than 3%.

A combination of dry and wet etching results in GaN nanowires whose diameter and density correspond to those of the original Pt nanoisland ensemble. Compared to prior studies on top-down GaN nanowires utilizing a metal dewetting approach, the nanowire ensembles fabricated in this work exhibit improvements in uniformity and a broader range of control over diameter and density with predictive power of the employed model. Although here GaN was chosen as the starting material for top-down fabrication, this approach can be easily transferred to other materials as the dewetting process takes place on a buffer layer. Therefore, this method is generally attractive for applications that require uniform nanowire ensembles over large areas.

### Acknowledgements

We are grateful to M. Mayer for inspiring discussions. We would like to thank S. Rauwerdink, W. Anders, and N. Volkmer for carrying out processing steps, A.-K. Bluhm for acquiring SE micrographs, and V.-D. Pham for critically reading the manuscript. The deposition of $SiN_x$ layers at Ferdinand-Braun-Institut, Leibniz-Institut für Höchstfrequenztechnik is gratefully acknowledged. This work has been financially supported by Deutsche Forschungsgemeinschaft under grant Ge2224/6-1.

**Supporting Information**

# Uniform large-area surface patterning achieved by metal dewetting for the top-down fabrication of GaN nanowires

Jingxuan Kang, Rose-Mary Jose, Miriam Oliva, Thomas Auzelle, Mikel Gómez Ruiz, Abbes Tahraoui, Jonas Lähnemann, Oliver Brandt, Lutz Geelhaar

*Paul-Drude-Institut für Festkörperelektronik, Leibniz-Institut im Forschungsverbund Berlin e.V., Hausvogteiplatz 5 – 7, 10117 Berlin, Germany*

1. **Procedure for statistical analysis of nanoisland size, shape, and number density**

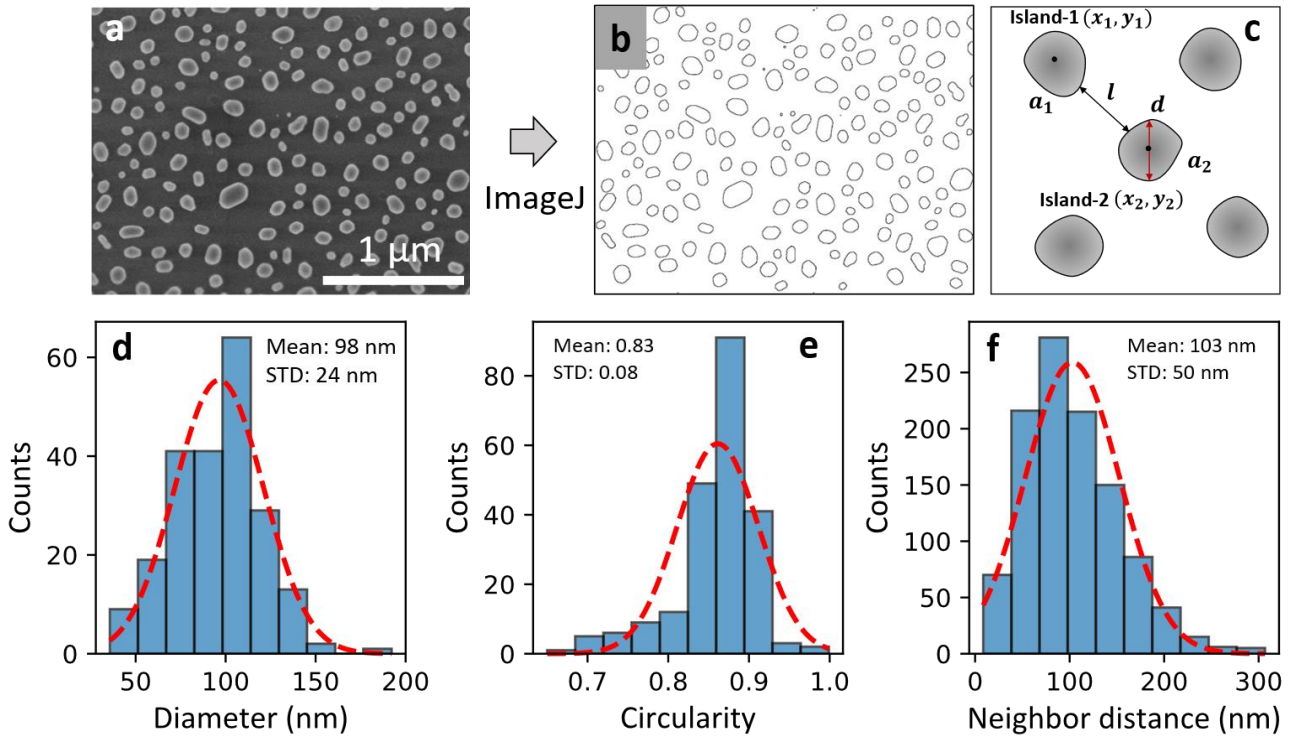

Figure S1: (a) SE micrograph obtained from a Pt dewetting sample with sample structure Pt(10 nm)/SiO$_x$(220 nm)/GaN, annealed for 1.5 h at 900 °C in an Ar-filled tube furnace. (b) Image after binarization by the software ImageJ. One SE micrograph with smaller magnification from the same sample is used for the subsequent statistical particle analysis. (c) Schematic illustration of the calculation of equivalent disc diameter and next-nearest-neighbour distance. Corresponding histograms of (d) equivalent disc diameter, (e) circularity, and (f) next-nearest-neighbour distance for the image in (b). The dashed curves are Gaussian fits, and the resulting mean and standard deviation (STD) are indicated in each diagram.

The statistical analysis of the nanoisland ensemble morphology is based on top-view secondary electron (SE) micrographs as shown in figure S1(a). First, the software ImageJ is used to identify the nanoislands by binarization of the brightness contrast compared to the substrate background as indicated in figure S1(b). Next, ImageJ provides for each Pt island the (*x, y*) coordinates of the center, the area *a*, and the perimeter *p*. Figure S1(c) illustrates how these data were processed by Python code to calculate the equivalent disc diameter *d* as well as the circularity *c* of each island and the next-nearest-neighbour distance *l* between two Pt islands by equations (1)–(3). In equation (3), ($x_1$, $y_1$) and ($x_2$, $y_2$) are the center coordinates and $d_1$ and $d_2$ the equivalent disc diameters of island-1 and island-2, respectively. Figures S1(d), (e), and (f) present histograms of island diameter, circularity, and next-nearest-neighbour distance extracted for an SE micrograph of the sample.

$$d = \frac{2\sqrt{a}}{\pi} \tag{1}$$

$$c = \frac{4\pi a}{p^2} \tag{2}$$

$$l = \sqrt{(x_1 - x_2)^2 + (y_1 - y_2)^2} - \frac{d_1}{2} - \frac{d_2}{2} \tag{3}$$

## 2. Histograms of nanoisland diameter and circularity

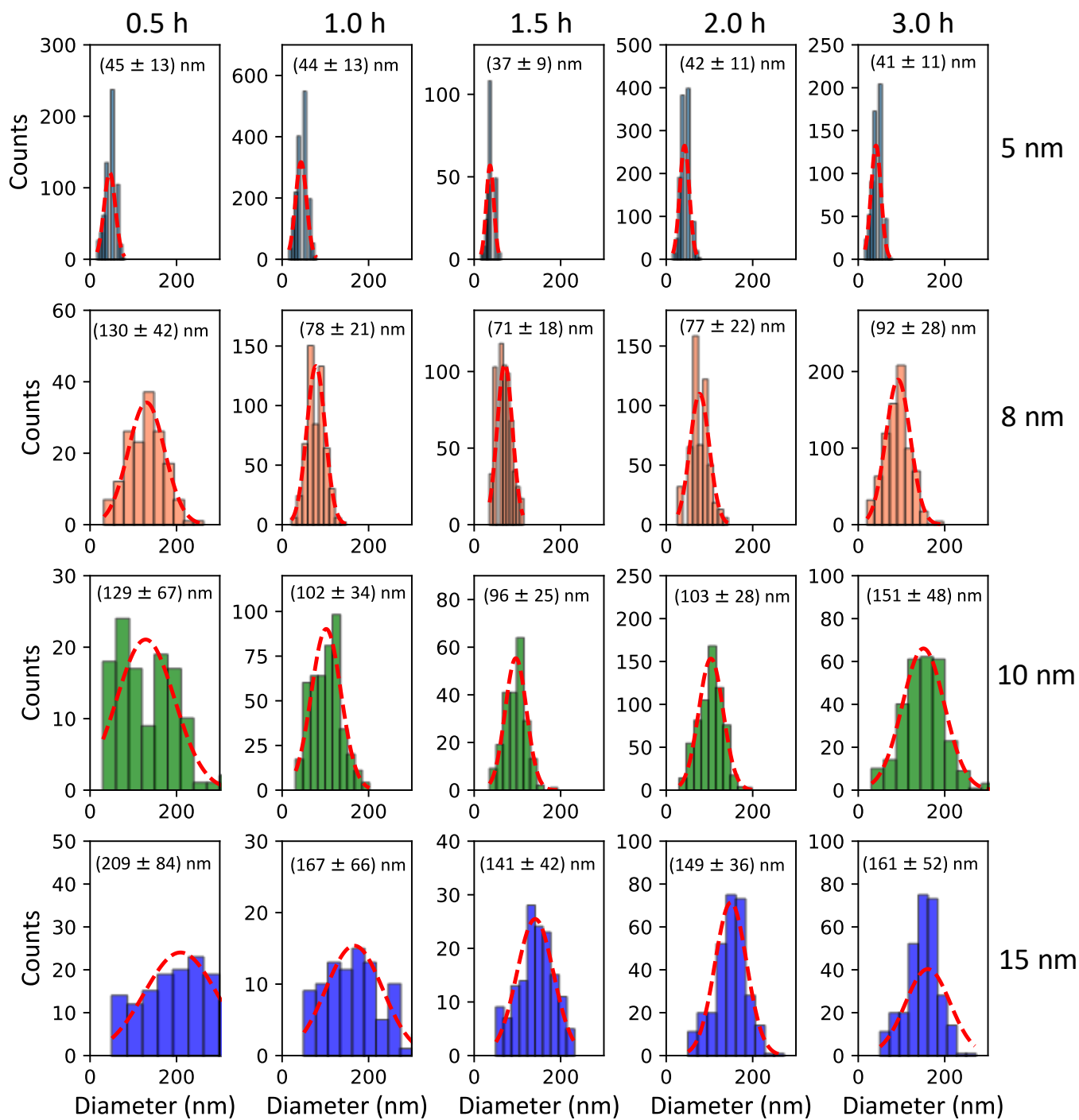

Figure S2: Histograms of Pt nanoisland diameter for dewetting samples with structure Pt/SiO$_x$(220 nm)/GaN and annealing temperature 900 °C. The annealing durations are indicated at the top of the figure and the Pt film thicknesses on the right-hand side. The dashed curves are Gaussian fits, and the resulting mean and STD are included in each diagram.

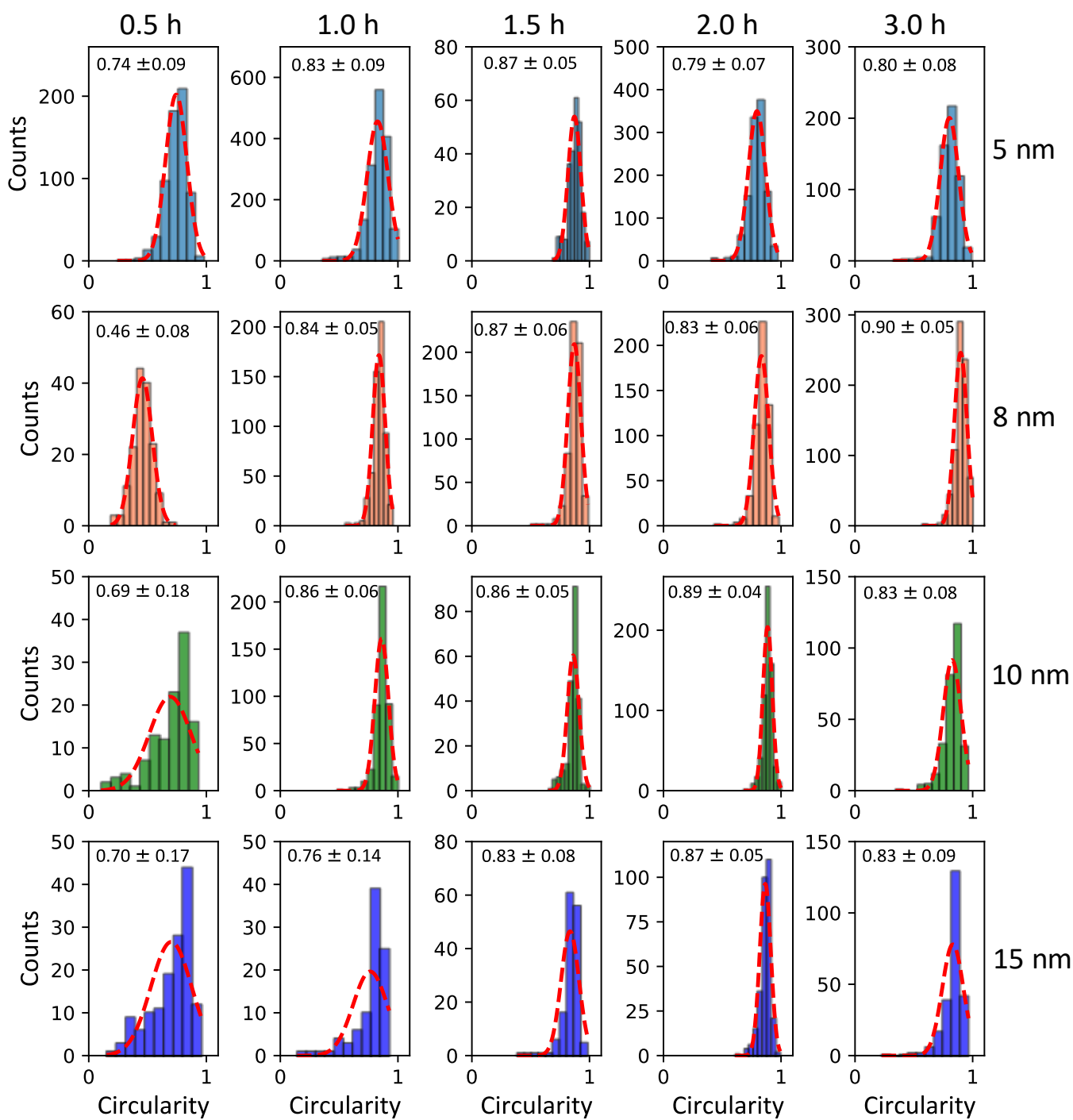

Figure S3: Histograms of Pt nanoisland circularity for dewetting samples with structure Pt/SiO$_x$(220 nm)/GaN and annealing temperature 900 °C. The annealing durations are indicated at the top of the figure and the Pt film thicknesses on the right-hand side. The dashed curves are Gaussian fits, and the resulting mean and STD are included in each diagram.

Figure S2 and S3 present histograms displaying the diameter and circularity distributions of Pt nanoislands, respectively. These data were obtained for samples with structure Pt/SiO$_x$(220 nm)/GaN with Pt film thicknesses ranging from 5 nm to 15 nm and annealing durations of 0.5 h to 3 h. The corresponding mean values of the distributions are shown in the main paper in figure 3(c) with an annealing temperature of 900 °C.

## 3. Nanoisland coalescence for very long annealing time

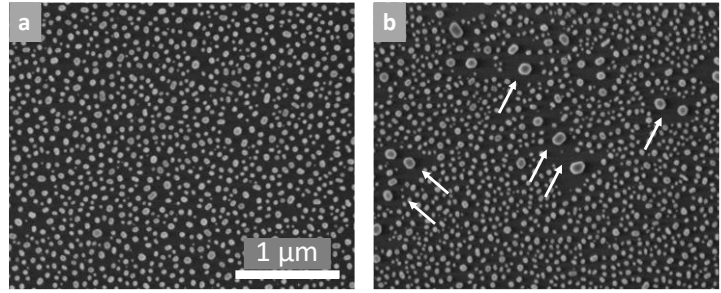

Figure S4: SE top-view micrographs of Pt(5 nm)/SiOx(220 nm)/GaN samples after (a) 1.5 h and (b) 3 h annealing at 900 °C in the tube furnace. The arrows in (b) indicate large islands formed by the coalescence of neighboring small islands.

As discussed in the main text and shown there in figure 3(c), the uniformity of the Pt nanoislands decreases after an optimal time $t_o$. The underlying mechanism is illustrated by the SE micrographs of Pt(5 nm)/SiO$_x$(220 nm)/GaN samples in figure S4. Panel (a) shows that the island size and shape distribution is uniform after annealing at 900 °C for 1.5 h. When the annealing duration is increased to 3 h, some islands appear that are clearly larger than the average, as depicted in figure S4(b). In other words, the size uniformity deteriorates. This is quantified by the statistical analysis whose results are presented in table S1. Around the large islands, the surface is depleted of other, smaller islands. Thus, the large islands form by the coalescence of islands in the homogenous ensemble. Therefore, this coalescence process limits the annealing duration that can be applied to improve the island size and shape homogeneity of the Pt nanoislands during dewetting.

Table S1: Statistical analysis of the samples in figure S4(a) and (b)

| Property | 1.5 h annealing | 3 h annealing |
|---|---|---|
| Mean diameter | 38 nm | 41 nm |
| Diameter standard deviation | 9 nm | 13 nm |
| Mean circularity | 0.87 | 0.80 |
| Circularity standard deviation | 0.02 | 0.09 |
| Density | 193 μm$^{-2}$ | 140 μm$^{-2}$ |

## 4. Pt diffusion during the annealing process

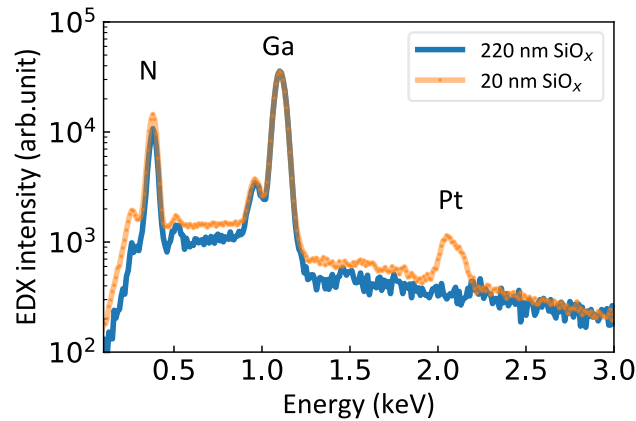

Figure S5: EDX spectra of GaN nanowire tops after removal of the SiO$_x$ layer. The two curves correspond to original SiO$_x$ buffer layer thicknesses of 20 nm (yellow) and 220 nm (blue).

The diffusion of Pt into the GaN substrate during annealing was studied for two samples with SiO$_x$ buffer layers between the Pt film and the GaN substrate. The buffer layer thicknesses were 20 nm and 220 nm. After annealing for 2 h at 900 °C and the etching procedure described in the main text, the buffer layers were removed in an HF bath and the top parts of the resulting

GaN nanowires were characterized by energy dispersive X-ray spectroscopy (EDX) in a scanning electron microscope (SEM). A Zeiss Ultra55 SEM operated at 10 kV and equipped with an EDAX Octane Elect Super EDX detector was used and the beam focused to the center of one of the larger nanowires. The spectra presented in figure S5 show the peaks associated with N and Ga, that are very similar in intensity for the two samples. In addition, only the sample with 20 nm buffer layer thickness exhibits a Pt peak . Hence, a $SiO_x$ buffer layer with a thickness of 220 nm effectively prevents the diffusion of Pt into GaN. Note that the detection limit of the system for Pt is between 0.05–0.1 atomic percent.